\documentclass[12pt, draftclsnofoot, onecolumn]{IEEEtran}
\usepackage[utf8]{inputenc}

\usepackage{amsmath}
\usepackage{amssymb}
\usepackage{amsfonts}
\usepackage{cite}

\usepackage{booktabs}
\usepackage{bbm}
\usepackage{mathrsfs}

\newcommand{\Rmnum}[1]{\expandafter\@slowromancap\romannumeral #1@}
\usepackage{graphicx}
\usepackage{subfigure}
\usepackage{epsfig}
\usepackage[numbers,sort&compress]{natbib}
\usepackage{xcolor}
\usepackage{color}
\usepackage{enumerate}
\usepackage{extarrows}
\usepackage[ruled,linesnumbered,lined]{algorithm2e}  
\usepackage{algorithmic}
\usepackage{tensor}
\usepackage[colorlinks,linkcolor=blue,anchorcolor=blue,citecolor=blue,urlcolor=black]{hyperref}
\usepackage{url}

\begin{document}

\title{Neuromorphic Wireless Cognition: Event-Driven Semantic Communications for Remote Inference}

\author{Jiechen Chen, \IEEEmembership{Student Member,~IEEE}, Nicolas Skatchkovsky,  \IEEEmembership{Member,~IEEE}, Osvaldo Simeone,~\IEEEmembership{Fellow,~IEEE}
\thanks{The work of Osvaldo Simeone and Nicolas Skatchkovsky was supported by the European Research Council (ERC) under the European Union’s Horizon 2020 research and innovation programme, grant agreement No. 725731, by an Open Fellowship of the EPSRC with reference EP/W024101/1, and by the European Union through project CENTRIC (101096379), and the work by Jiechen Chen was funded by the China Scholarship Council and King’s College London for their Joint Full-Scholarship (K-CSC) under Grant CSC202108440223.}
\thanks{The authors are with the King’s Communications, Learning and Information Processing (KCLIP) lab, King’s College London, London, WC2R 2LS, UK. (email:\{jiechen.chen, nicolas.skatchkovsky, osvaldo.simeone\}@kcl.ac.uk).}
}

\maketitle
\vspace{-1.5cm}

\begin{abstract}
Neuromorphic computing is an emerging computing paradigm that moves away from batched processing towards the online, event-driven, processing of streaming data. Neuromorphic chips, when coupled with spike-based sensors, can inherently adapt to the ``semantics'' of the data distribution by consuming energy only when relevant events are recorded in the timing of spikes and by proving a low-latency response to changing conditions in the environment. This paper proposes an end-to-end design for a neuromorphic wireless Internet-of-Things system that integrates spike-based sensing, processing, and communication. In the proposed NeuroComm system, each sensing device is equipped with a neuromorphic sensor,  a spiking neural network (SNN), and an impulse radio (IR) transmitter with multiple antennas. Transmission takes place over a shared fading channel to a receiver equipped with a multi-antenna impulse radio receiver and with an SNN. In order to enable adaptation of the receiver to the fading channel conditions, we introduce a hypernetwork to control the weights of the decoding SNN using pilots. Pilots, encoding SNNs, decoding SNN, and hypernetwork are jointly trained across multiple channel realizations. The proposed system is shown to significantly improve over conventional frame-based digital solutions, as well as over alternative non-adaptive training methods, in terms of time-to-accuracy and energy consumption metrics. 
\end{abstract}

\begin{IEEEkeywords}
\vspace{-0.2cm}
Neuromorphic computing, spiking neural networks, semantic communications.
\end{IEEEkeywords}

\IEEEpeerreviewmaketitle
\newtheorem{definition}{\underline{Definition}}[section]
\newtheorem{fact}{Fact}
\newtheorem{assumption}{Assumption}
\newtheorem{theorem}{\underline{Theorem}}[section]
\newtheorem{lemma}{\underline{Lemma}}[section]
\newtheorem{proposition}{\underline{Proposition}}[section]
\newtheorem{corollary}[proposition]{\underline{Corollary}}
\newtheorem{example}{\underline{Example}}[section]
\newtheorem{remark}{\underline{Remark}}[section]
\newcommand{\mv}[1]{\mbox{\boldmath{$ #1 $}}}
\newcommand{\mb}[1]{\mathbb{#1}}
\newcommand{\Myfrac}[2]{\ensuremath{#1\mathord{\left/\right.\kern-\nulldelimiterspace}#2}}
\newcommand\Perms[2]{\tensor[^{#2}]P{_{#1}}}
\newcommand{\note}[1]{[\textcolor{red}{\textit{#1}}]}

\section{Introduction}\label{sec:intro}
\subsection{Context and Motivation}
The recent rollout of 5G around the world has marked the start of a switch of telecom systems from \emph{network-centric carriers of bits} to \emph{user-centric distributed processors of intelligence}. A key element of this switch will be the integration of wireless systems with sensing and cognition, a technology trend we will refer to as \emph{wireless cognition}. In this context,  this paper is motivated by the two following paradigm shifts that are widely envisioned as central to 6G: \\ \noindent $\bullet$ \emph{From universality to goal-driven specialization:} With the emergence of wireless cognition, there is a need to develop semantics-aware solutions that integrate sensing, communication, and computing, tailoring resource consumption to the goals of the task at hand \cite{popovski2020semantic, strinati20216g, kountouris2021semantics, kalfa2021towards, qin2021semantic}.
\\
  \noindent $\bullet$ \emph{From hardware agnostic to hardware-constrained design:} By assuming universal computing architectures, the conventional design  of communication systems is agnostic to the specific hardware systems deployed at the communicating nodes. This approach fails to acknowledge the critical importance of the computing architecture in the effective and efficient implementation of semantic tasks (see, e.g., \cite{conti2018xnor}). 
  
  %For example, BR's prior work has shown that implementing AI inference tasks on optimised  memristive analog accelerators can result in $88\times$ speed up compared to digital CMOS designs \cite{Rajendran75}.
  
  %... consumes while... {\color{red} can we give a simple example comparing CPU processing to GPU processing, say, in terms of time and power? even better if we can cite some of your work} While the GPU ... is still prohibitively high for mobile applications, it is significantly lower than...., demonstrating the importance of hardware co-design.
  
  %several \textbf{separation principles: memory from processing,} with specialised cognitive tasks involving  the extraction and exploitation of meaning from data and as Moore's law slows down, this approach, while highly dependable, is showing its limitations. 
   \emph{Neuromorphic sensing and computing} are emerging as alternative, \emph{brain-inspired}, paradigms for efficient data collection and semantic signal processing. The main features of the technology are energy efficiency, native event-driven processing of time-varying semantic sources, spike-based computing, and always-on on-hardware adaptation \cite{mehonic2022brain,davies2021advancing}. 
   
   \noindent $\bullet$ \emph{Neuromorphic sensors} encode information in the timing of spikes, and include neuromorphic cameras, silicon cochleas, and brain-computer interfaces. As a general principle of operation, spikes are produced only when relevant changes occur in the signals being sensed \cite{inivi,event,rebecq2019events,lee2019neuro}. 
   
   \noindent $\bullet$ \emph{Neuromorphic processors}, also known as \emph{spiking neural networks (SNNs)}, are networks of dynamic spiking neurons that mimic the operation of biological neurons \cite{neuro}. Spiking neurons communicate and process with the timings of spikes \cite{jang2019introduction}. When implemented on specialized -- digital or mixed analog-digital -- hardware or on tailored FPGA configurations, SNNs have minimal idle and operating energy cost, and consume as little as \emph{a few picojoules} per spike \cite{rajendran2019low}.

 Current commercial use cases of neuromorphic technologies range from drone monitoring via  Dynamic Vision Sensor (DVS) cameras \cite{hu2016dvs,lichtsteiner2006128} through the development of brain-computer interfaces\footnote{\url{https://neuralink.com/}} to the development of fast and accurate COVID-19 antibody testing\footnote{\url{https://bit.ly/3r0NeoS}}. Neuromorphic computing platforms include Intel’s Loihi SNN chip, IBM’s TrueNorth, and Brainchip’s Akida \cite{davies2018loihi, win1998impulse}. This work views the emergence of neuromorphic technologies as a unique opportunity for the development of \emph{goal-driven, specialized, and hardware-constrained wireless cognition}.

 To date, work on the integration of wireless connectivity and neuromorphic systems has been very limited, including only a specific implementation introduced for biomedical applications in  \cite{shahshahani2015all}.

\subsection{NeuroComm}

\begin{figure}[htp]
	\centering
	\includegraphics[width=5in]{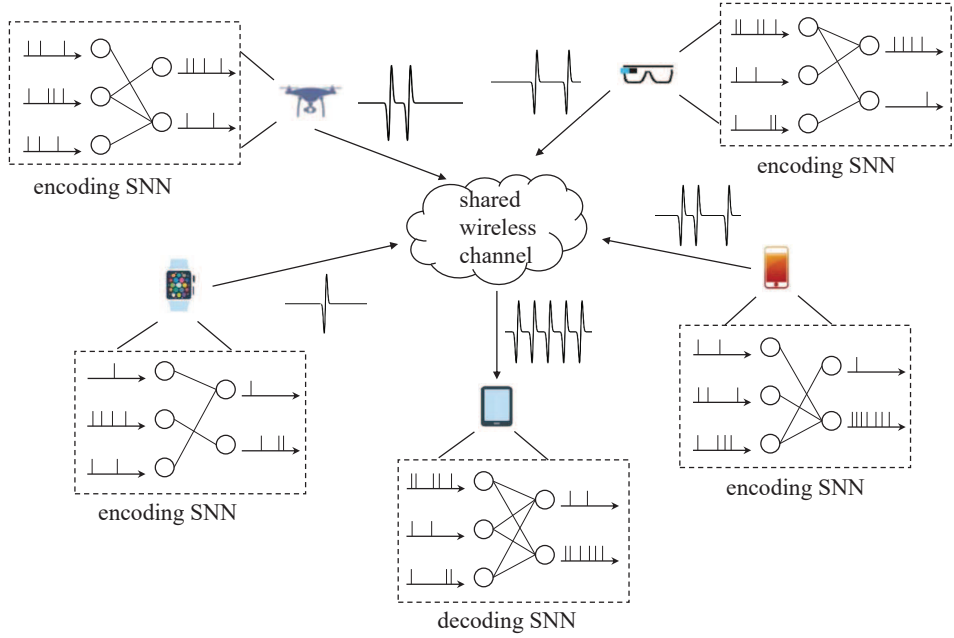}
	\caption{This paper focuses on an edge-based IoT system in which data are collected and pre-processed in a streaming fashion by mobile sensors so that inference can be carried out continuously at another edge device. The proposed NeuroComm solution integrates neuromorphic sensing, neuromorphic computing -- via spiking neural networks (SNNs) -- and impulse radio. \label{gmodel}}
	
\end{figure}

This paper introduces \emph{NeuroComm}, a wireless cognition system that integrates neuromorphic sensing, computing, and spike-based communications.  As illustrated in Fig~\ref{gmodel}, NeuroComm consists of a wireless Internet-of-Things (IoT) network in which multiple wireless sensing devices communicate over a shared wireless link to an edge device. We target an \emph{always-on} monitoring application in which the environment being observed by the sensors presents \emph{sparse} activity, and \emph{inference} at the receiving edge device should be carried out in a \emph{timely} manner. 

An example of the scenarios under consideration is given by distributed sensors deployed to monitor the presence of drone activity, e.g., in an airport. In such applications, sensors capture uninformative signals most of the time, while needing to promptly enable object recognition and detection at the receiving processor when relevant activity is present in the system. 

\begin{figure}[!htb]
	\begin{center}
		\subfigure[Digital communication system.{\label{rep1}}]{\includegraphics[width=6.2in]{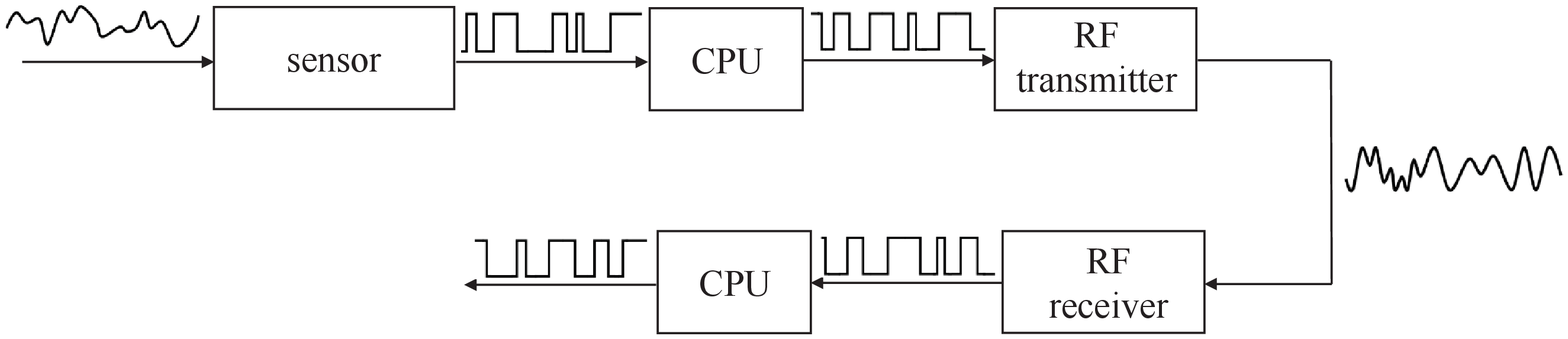}}\hfill
		\subfigure[Neuromorphic communication system.{\label{rep2}}]{\includegraphics[width=6.2in]{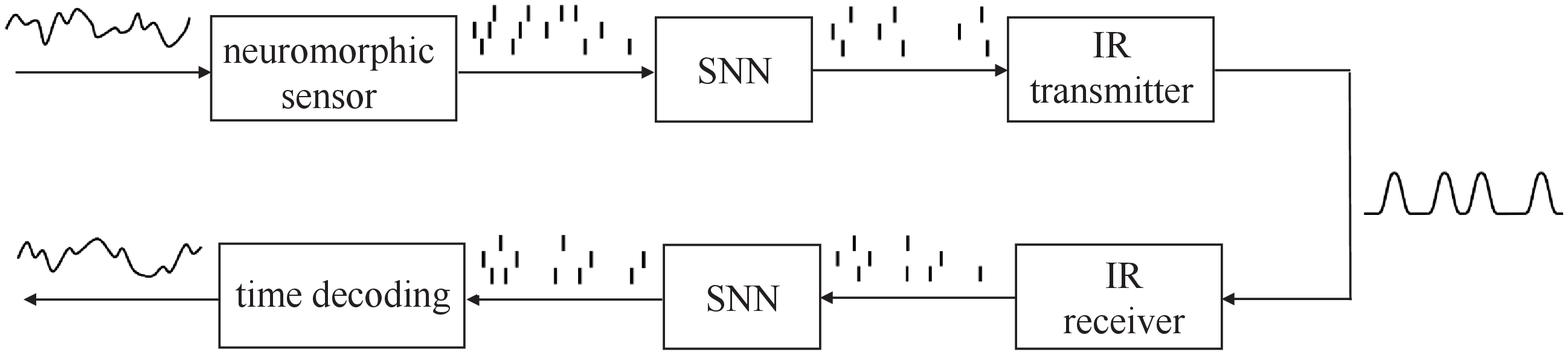}}
	\end{center}
	\caption{(a) Illustration of a conventional digital implementation of an IoT link with remote inference, in which sensed information is transformed into bits, packaged into frames, encoded, and transmitted; (b) Illustration of the proposed NeuroComm system in which digital blocks are replaced by neuromorphic blocks, including neuromorphic sensing and processing via SNNs. In NeuroComm, spikes are communicated by means of impulse radio \textcolor{blue}{(IR)} and processed via SNNs. \label{replace}}
\end{figure}

As depicted in Fig~\ref{replace}-(a), a conventional deployment for remote inference would rely on  digital \emph{frame-based} sensing, processing, and computing blocks. This approach has two major drawbacks: (\emph{i})  \emph{Semantic-agnostic energy consumption}:  Conventional digital sensing devices would generally sense and transmit frames continuously, irrespective of the level of ``information'' present in the sensed signal for the given application; (\emph{ii}) \emph{Latency quantization}: Sensed information would be packaged into frames, enabling the receiver to enhance its inference accuracy only upon the reception and processing of entire frames. 

 In NeuroComm, as illustrated in Fig~\ref{replace}-(b), we propose to replace the digital frame-based blocks of the conventional deployment in  Fig~\ref{replace}-(a) with neuromorphic sensing, computing, and transmission blocks. The main goals of these system-level innovations are as follows: (\emph{i})  \emph{Semantic-aware energy consumption}:   First, we wish to ensure an energy consumption level throughout the sensing-processing-communication chain that adapts to the dynamics of the environment being monitored; (\emph{ii}) \emph{Enhanced time-to-efficiency}: Second, we aim at enabling a graceful improvement in the inference performance over time as more information on the environment is sensed and communicated. 
 
 These goals are met by leveraging the \emph{event-driven} nature of neuromorphic sensing and computing, as well as the synergy between spike processing and pulse-based transmission via \emph{impulse radio} \cite{win2000ultra}. Neuromorphic sensors, SNNs, and impulse radio blocks consume energy only when spikes are produced, reflecting patterns of activity in the monitored scene. The use of impulse radio is also appealing in light of the role of wideband transmission for THz communications in 6G, as well as in existing low-power communication standards like IEEE 802.15.4 \cite{akyildiz2022terahertz, yang20196g}. 
 
 %Spiking neural networks, which are networks of dynamic spiking neurons, operate over time by replacing the real-valued outputs of traditional artificial neural networks (ANNs) with binary spikes. Finally, impulse radio can be employed to encode spikes into pulses for communication. 

% In order to leverage the sparsity of the activity of the environment, i.e., the semantics of the  application, we assume that the devices are equipped with neuromorphic sensors, such as DVS cameras.  goal is to implement continuous remote inference via distributed sensors and an edge processor in an energy-efficient manner.  The reference setting of interest is 

\subsection{Main Contributions}

\begin{figure}[htp]
	\centering
	\includegraphics[width=6.2in]{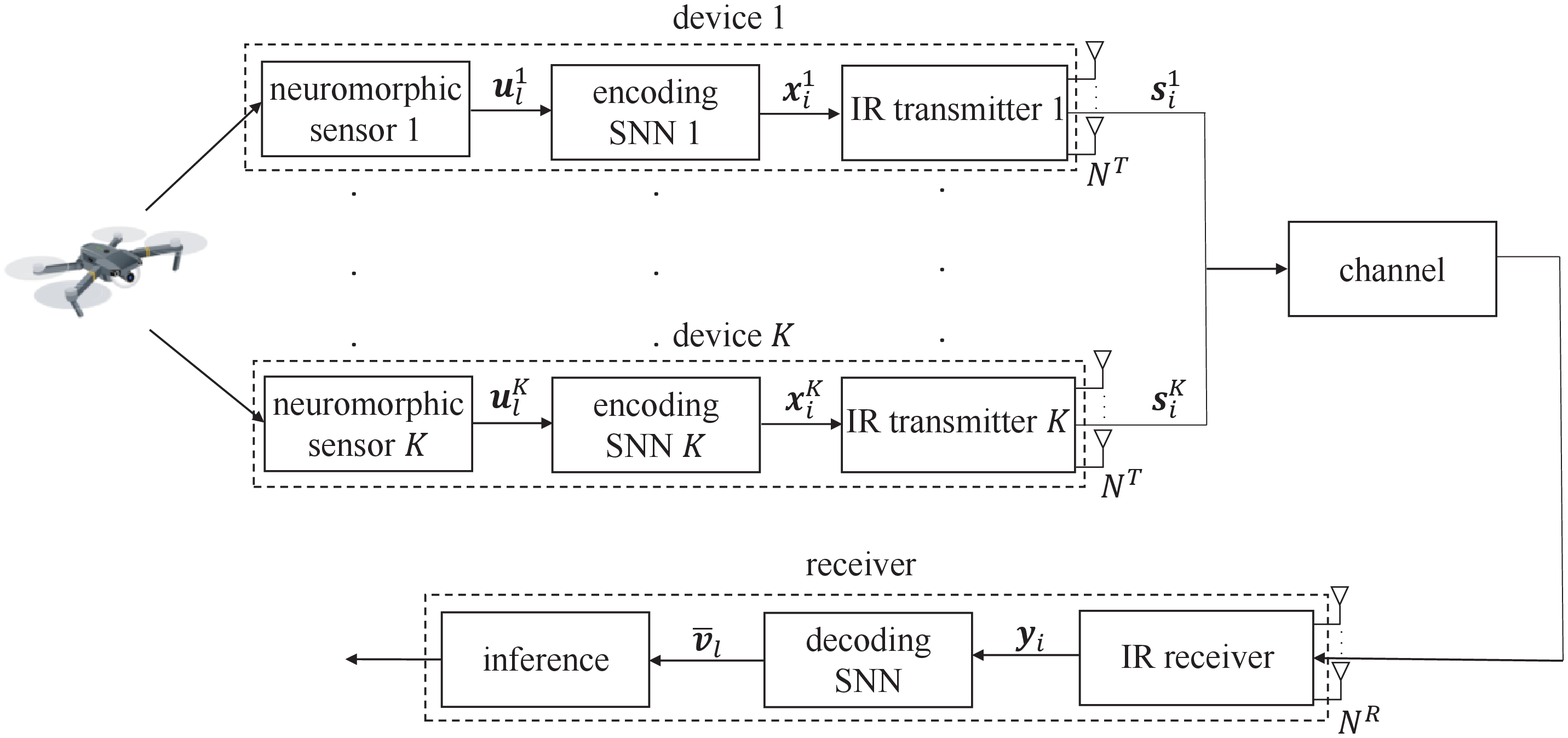}
	\caption{A block diagram illustrating the main operations involved in the implementation of the proposed NeuroComm system (IR stands for impulse radio). }
	\label{model}
\end{figure}

As depicted in Fig. \ref{model}, NeuroComm implements a novel form of \emph{joint source-channel coding} whereby data from neuromorphic sensors are pre-processed by SNNs and directly mapped into electromagnetic pulses on shared wireless channels in an online, streaming, fashion. The receiver processes the received signals via an SNN to carry out inference. This approach differs from the standard \emph{separate source-channel coding} solution illustrated in Fig. \ref{modelconv} that is implemented by digital frame-based systems. In the conventional system, data from sensors are first compressed and then encoded on a frame-by-frame basis.

The main contributions of this paper are summarized as follows. \begin{itemize}
\item We introduce NeuroComm, a novel architecture for wireless cognition integrating neuromorphic sensing, processing, and communications.
\item We propose an end-to-end design of the NeuroComm architecture based on supervised learning via surrogate gradient descent methods.
\item In order to enable adaptation to channel conditions,  we introduce a hypernetwork-based approach. The hypernetwork is trained to take as input signals received from pilot symbols to control the weights applied by the receiving SNN. 
\item Extensive numerical results are provided that demonstrate the advantages of the proposed architecture and design over conventional frame-based digital solutions, as well as over alternative non-adaptive training methods, in terms of time-to-accuracy and energy consumption metrics. 
\end{itemize}

\subsection{Related Work}
To the best of our knowledge, the only paper that has proposed the idea of end-to-end learning for a communication link with neuromorphic encoder and decoder is the conference publication \cite{skatchkovsky2020end} by a subset of the authors of this paper. This is a preliminary study focusing on a special case of the NeuroComm system in which a single transmitter operates over a frequency-flat channel with input data collected with a neuromorphic camera. In contrast, the current work investigates the general case of multiple devices communicating over a shared fading channel.  

A paper \cite{borsos2022resilience} by a team at Ericsson has considered the distributed implementation of an SNN on a packet-based wireless channel, in which the SNN is trained  offline in a channel-agnostic manner. Moreover, a recent work \cite{equ} proposed a decision-feedback equalizer based on an SNN as a solution for an energy-efficient receiver.

There is a rich literature on joint source-channel coding for conventional artificial neural networks (ANNs). For example, \cite{bourtsoulatze2019deep, yang2021deep, kurka2021bandwidth}  investigated wireless image transmission based on convolutional neural networks. As another example, the authors in \cite{alawad2021end} optimized intelligent reflecting surfaces-assisted wireless communication systems in an end-to-end manner by treating the transmission chain as a joint source-channel coding problem. 

As for the applications of hypernetworks in communications, reference \cite{goutay2020deep} studied an application to MIMO detection under highly correlated channels to avoid retraining for each channel realization. The paper \cite{liu2021hyperrnn} proposed a hypernetwork-based architecture for recurrent neural networks that leverages partial uplink-downlink reciprocity and temporal correlation of the fading channel to perform channel estimation in frequency-division duplex systems. The use of hypernetworks coupled with SNNs appears to be novel.

\subsection{Organization}

The remainder of the paper is organized as follows. Section \ref{sec:System Model} presents the system model for multi-device remote inference. The conventional frame-based digital transmission is reviewed for reference in Section \ref{conventional method}. The NeuroComm architecture is proposed in Section \ref{neuro}, while Section \ref{sec:training} presents the integration of NeuroComm with a hypernetwork and the resulting training problem. Experimental setting and results are described in Section \ref{exp}. Finally, Section \ref{con} concludes the paper.

\section{System Model}\label{sec:System Model}
In this paper, we study the wireless edge-based IoT system illustrated in Fig.~\ref{gmodel}, whose goal is to implement continuous remote inference via distributed sensors and an edge processor in an energy-efficient and timely manner. To this end, the system integrates neuromorphic computing via spiking neural networks (SNNs) and impulse radio (IR) transmission. In this section, we formally describe setting, channel model, the operation of neuromorphic sensors, and the transmission model.

\subsection{Setting}

As illustrated in Fig.~\ref{gmodel}, we consider a remote inference system with one receiver collecting information from $K$ sensing devices. Each device $k\in\mathcal{K}=\{1,\ldots,K\}$ is equipped with a neuromorphic sensor and with $N^T$ antennas, while the receiver has $N^R$ antennas. Communication takes place over a shared multi-path fading channel under average or peak power constraints. 

A conventional digital sensor, such as a digital camera, would capture natural signals via sampling and  analog-to-digital conversion. Accordingly, in a continuous, always-on, monitoring application, a digital sensor would produce a constant stream of bits, typically packaged into frames. In this paper, in order to leverage efficiencies related to the assumed sparsity of the dynamics of the environment being monitored, we assume that sensing devices are equipped with neuromorphic sensors, such as DVS cameras \cite{skatchkovsky2021learning}. We recall that, by the principle of operation of a neuromorphic sensor, the presence of a spike generally indicates the occurrence of an event. For instance, with DVS cameras, a spike is produced as the result of a sufficiently large change in brightness at a pixel \cite{lichtsteiner2006128}. 

We adopt a discrete-time model, with discrete time index $l=1,2,...,L$ used to denote the time samples produced by the neuromorphic sensors. We assume the devices to be time synchronized.  The sensor available at each device produces $D_u$ spiking signals. For example, for a DVS camera, $D_u$ is the number of pixels. Accordingly, the sensor at the $k$th device outputs a $D_u \times 1$ vector $\mv u_l^k$ with binary entries at each discrete time $l$. Each binary entry $u_{l,m}^k$ of the $D_u\times1$ vector $\mv u_{l}^k=[u_{l,1}^k,\dots,u_{l,D_u}^{k}]\in\{0,1\}^{D_u}$ represents the presence $(u_{l,m}^k=1)$, or the absence $(u_{l,m}^k=0)$ of a spike in the $m$th spiking signal at time $l$. As mentioned, a spike may indicate that the corresponding $m$th pixel has undergone a sufficiently large change in luminance during the $l$th time interval.

The signals $\mv u_l^k$ captured by each $k$th neuromorphic sensor are processed and transmitted over a fading multiple access channel to a receiver. The receiver's goal is to carry out an inference task, such as classification, based on the received signals. In the next two sections, we will first consider for reference a conventional digital communication and processing chain depicted in Fig.~\ref{frame}, and we will then introduce the proposed NeuroComm system, illustrated in Fig. \ref{model}.

\subsection{Channel Model}
We assume a general multi-path propagation model, which may account, e.g., for propagation in the sub-6GHz band, as in the IEEE 802.15.4z standard, or for millimeter wave and Terahertz (THz) bands \cite{han2015multi}. Accordingly, the continuous-time channel response between the $m$th antenna of device $k$ and the $n$th antenna of the receiver can be expressed as 
\begin{align}
	h^k_{n,m}(t)=\sum_{p=1}^{N_P} \beta^k_{n,m,p} g(t-\tau^k_{n,m,p}),
	\label{h}
\end{align}
where $N_{P}$ is the total number of resolvable propagation paths, including possibly a line-of-sight (LoS) component;  $\beta^k_{n,m,p}$ and $\tau^k_{n,m,p}$ represent the complex amplitude and delay of the $p$th path, respectively; and $g(\cdot)$ represents the matched filter response (see, e.g., \cite{jacobs1965principles}). The maximum delay of all channels is denoted as $T_h$, which is measured in seconds, i.e., we have $h_{n,m}^k(t)=0$ for $t\geq T_h$ and for all values of $n$, $m$, and $k$. The channel is assumed to be constant for the duration of a transmission frame. A frame is assumed to be sufficiently long to encompass the presentation of input signals $\mv u_l^k$ of $L$ samples, $l=1,\ldots,L$, from the neuromorphic sensor to each device $k$.

\subsection{Transmission Model}
The modulated signal emitted by the $m$th antenna of the impulse radio transmitter $k$ is denoted as $s_m^k(t)$ and is assumed to have bandwidth $1/T_c$, where $T_c$ is measured in seconds. We write the overall duration of a transmission frame as $L L_b T_c$, where integer $L_b$ is referred to as the \emph{bandwidth expansion factor}. Since $L$ is the number of samples produced by the sensors per frame, when sampled at Nyquist rate $1/T_c$, the transmitted and received signals present $L_b$ samples for each sample of the sensed signals. 

The signal obtained by the $n$th antenna of the receiver is given by the superposition of the signals transmitted by all the $N^T$ antennas at the $K$ devices as
\begin{align}
	y_n(t)=\sum_{k=1}^{K} \sum_{m=1}^{N^T} s^k_m(t)*h^k_{n,m}(t)+z_n(t),
	\label{re}
\end{align}
where $z_n(t)$ is the additive white Gaussian noise with power spectral density $N_0$; ``$*$'' denotes the convolution operation; and $h^k_{n,m}(t)$ is the (complex) channel input response in \eqref{h} between antenna $m$ of device $k$ and receive antenna $n$.

Sampling signal \eqref{re} at rate $1/T_c$ yields the discrete-time signal $y_{i,n}=y_n(iT_c)$, for $i=1,2,\ldots,LL_b$, across the time frame. Having a maximum delay spread of $T_h$, upon sampling, the effective discrete-time channels have $L_h=T_h/T_c$ taps. Accordingly, the channel between antenna $m$ of device $k$ and antenna $n$ of the receiver is described by the $L_h\times1$ vector $\mv h_{n,m}^k=[h_{n,m}^k(0), h_{n,m}^k(T_c),\ldots,h_{n,m}^k((L_h-1)T_c)]^T$. Therefore, the received discrete-time signal $y_{i,n}$ at the $n$th receive antenna can be expressed as
\begin{align}
	 y_{i,n}= \sum_{k=1}^K \sum_{m=1}^{N^T} (\mv h_{n,m}^k)^T \mv s_{i,m}^k + z_{i,n},
	\label{y_l}
\end{align}
where we have defined the $L_h \times 1$ vector $\mv s_{i,m}^k=[s_{i,m}^k, s_{i-1,m}^k, \ldots, s_{i-L_h+1,m}^k]^T$ with $s_{i,m}^k=s_m^k(iT_c)$, and $s_{i,m}^k=0$ if $i\leq0$; and $z_{i,n}=z_n(iT_c)\sim \mathcal{CN}(0,N_0)$. 

We consider two different types of constraints on the transmitted signals under the assumption that the transmitters are not aware of the current channel realization.
\begin{itemize}
\item \emph{Per-frame energy constraint:} Collecting the transmitted samples from all the $N^T$ antennas of device $k$ in a frame in the $LL_b N^T \times 1$ vector $\mv s^k$, the per-frame energy constraint imposes the inequality as
\begin{align}
	 \|\mv s^k\|^2 \leq E^k_{\rm fr},
	\label{pc}
\end{align}
where $E^k_{\rm fr}$ is the maximal per frame transmission energy for transmitter $k$. 
\item \emph{Per-symbol energy constraint:} The per-frame energy constraint \eqref{pc} is well suited for conventional frame-based system, and is not a natural choice for systems operating in a streaming fashion. Therefore, we also consider the per-symbol energy constraint as
\begin{align}
	 |s_{i,m}^k|^2\leq E^k_{\rm s},
	\label{sc}
\end{align}
where $E^k_{\rm s}$ is the per-symbol energy constraint for transmitter $k$. 
\end{itemize}

\subsection{Pilot Transmission}
In order to enable adaptation of the receiver to changing channel conditions, we assume that pilot symbols are transmitted by each device prior to the start of data transmission in each frame. Pilot transmissions from different devices are assumed to be orthogonal. Accordingly, device $k$ transmits $L_\textrm{p}$ symbols at rate $1/T_c$ from each antenna. Writing as $s_{i,m}^k$ with $i=-L_{\textrm{p}}+1, -L_{\textrm{p}}+2, \ldots, 0$, as the pilot symbols sent by device $k$ from antenna $m$, the signal received from device $k$ during the training phase can be written as 
\begin{align}
	 y_{i,n}^k= \sum_{m=1}^{N^T} (\mv h_{n,m}^k)^T \mv s_{i, m}^k + z_{i,n},
	\label{y_p}
\end{align}
where we have defined the $L_h \times 1$ vector $\mv s_{i,m}^k=[s^k_{-L_{\rm p}+1+i,m},s^k_{-L_{\rm p}+i,m},\ldots, s^k_{-L_{\rm p}-L_h+i+2,m}]^T$, and $s_{i,m}^k=0$ if $i\leq-L_{\rm p}$. We impose the energy constraint
\begin{align}
	 \|\mv s_{\textrm{p}}^k\|^2 \leq E^k_{\rm p},
	\label{pilotc}
\end{align}
which applies to the vector $\mv s_{\textrm{p}}^k$ of all pilot symbols $s_{i,m}^k$ transmitted at time instants $i=-L_{\rm p}+1,\ldots,0$ from all antennas $m=1,...,N^T$, and $E_{\rm p}^k$ is the maximum pilot energy constraint for transmitter $k$.

\begin{figure}[htp]
	\centering
	\includegraphics[width=6.2in]{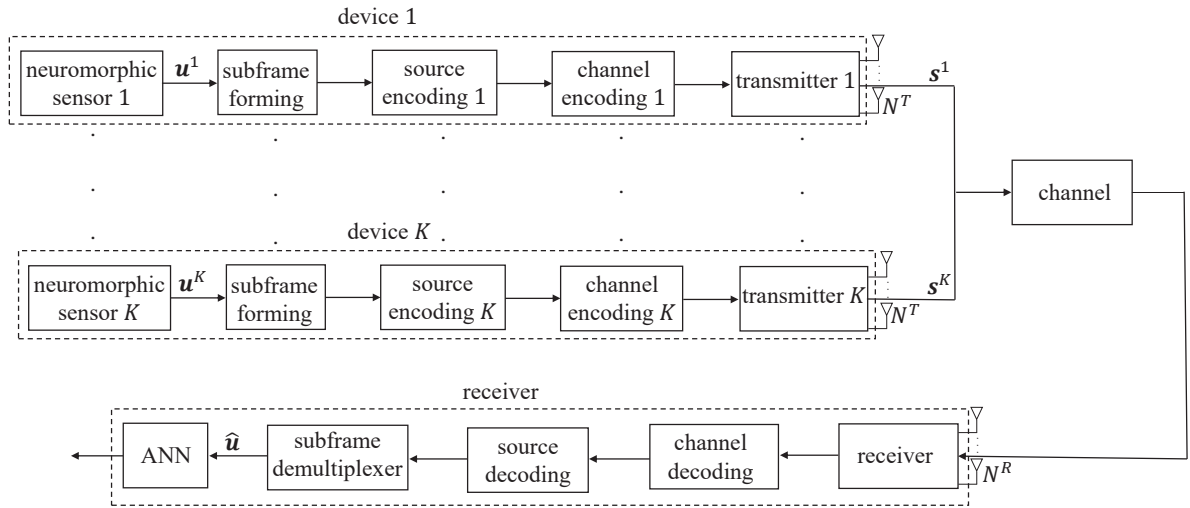}
	\caption{Block diagram illustrating the main operations involved in a standard digital frame-based wireless remote inference system with neuromorphic sensing. \label{modelconv}}
	\label{frame}
\end{figure}

\section{Conventional Frame-Based Digital Transmission} \label{conventional method}
 
For reference, in this section we review the conventional frame-based digital system illustrated in Fig. \ref{modelconv}. As discussed in Sec. \ref{sec:intro}, this solution implements \emph{separate source-channel coding} by processing the signals produced by the sensors via frame-based source and channel encoding and decoding blocks at the transmitters and at the receiver. While the sensors are neuromorphic for consistency with the setting under study in this work, unlike in the proposed NeuroComm architecture, all other blocks adopt standard digital computing and communication technologies. 

To elaborate, as shown in Fig.~\ref{frame}, at each device $k$, the samples $\mv u^k$ produced by the neuromorphic sensor are first divided into \emph{subframes}. Each subframe is compressed via a source encoder, and then channel encoded. The channel encoded sequence is transmitted using standard digital modulation on the shared channel to the receiver. The receiver carries out channel decoding for the signals sent by all devices. The decoded signals are decompressed via the source decoder to recover an estimate of the original subframes produced by the sensors. These estimates are demultiplexed to produce an estimate of the sensed signals $\mv u^k$ with $k=1,\ldots,K$, and then used for inference. Inference is carried out by a standard ANN at the receiver.

\subsection{Separate Source-Channel Coding}

The samples $\mv u_l^k$ produced by each $k$th neuromorphic sensor across time instant $l=1,....,L$ are formatted into subframes of $L_{\rm enc}$ samples each. As a result, $F=\lceil L/L_{\rm enc}\rceil$ subframes $\mv u_f^k$ with $f=1,\ldots,F$ are produced, with each subframe $\mv u_f^k$ containing $D_u L_{\rm enc}$ binary digits. Note that if the ratio $L/L_{\rm enc}$ is not an integer, the last subframe can be padded with zeros.

Each subframe $f=1,\ldots,F$ is compressed, producing a shorter binary sequence, which is then channel encoded, obtaining $L_b L_{\rm enc}$ samples $s^k_{i,m}$, with $i=(f-1)L_bL_{\rm enc}+1,\ldots, fL_bL_{\rm enc}$, for all transmit antennas $m=1,\ldots,N^T$. Recall that the bandwidth expansion factor $L_b$ specifies the number of channel uses per sensed sample that are available on the communication channel. The choice of the subframe size $L_{\rm enc}$ generally entails a trade-off between latency due to subframe processing and communications, on the one hand, and the effectiveness of block encoding, on the other hand, which is known to improve as the blocklength increases \cite{thomas2006elements}.

Each discrete-time symbol in the sequence $s_{i,m}^k$ is modulated using a waveform of bandwidth $1/T_c$, and transmitted over the wireless shared channel \eqref{h} to the receiver. Based on the received signal \eqref{y_l} in each subframe $f$, the receiver performs channel decoding and source decoding to recover an estimate $\hat{\mv u}^k_f$ of the original $f$th subframe. A specific implementation of source and channel encoders/decoder is described in Section \ref{exp}.

\subsection{Inference}
While the proposed framework can be applied more broadly, in this work, we focus on supervised learning for a classification. In this subsection, we describe how inference is implemented, while the next subsection covers training.

The inference task amounts to a mapping between the sensed input $\mv u^k$ from all devices $k=1,\ldots,K$ and a target variable $\mv v$. Given our focus on classification, we consider $\mv v$ to be a $D_v \times 1$ one-hot vector representing the class index, where $D_v$ is the number of classes. For the conventional system under study, we introduce an ANN that takes as input the per-subframe estimated  sensed signals $\hat{\mv u}^k_f$ produced by the source decoder for all $k=1,\ldots,K$ to output an estimate $\hat{\mv v}_f$ of the target vector. The decisions produced for each subframe $f$ are then combined to output a final decision $\hat{\mv v}$, e.g., via a majority rule. This approach is useful to enable the ANN to make a decision after the reception of each subframe. Other conventional solutions may involve the use of Recurrent Neural Networks (RNNs), and they are not further considered here.

Mathematically, the ANN defines a parametric mapping $\hat{\mv v}_f=G_{\scriptsize \mv \phi}(\hat{\mv u}_f)$, where $\mv \phi$ is the vector of parameters of the ANN, and the output $\hat{\mv v}_f\in[0,1]^{D_v}$ provides an estimate of the probabilities of the $D_v$ classes, which is obtained via a softmax output layer. The decision at the $f$ subframe can be made by choosing the class with the largest probability in vector $\hat{\mv v}_f$. After each subframe $f$ is received, a more accurate decision $\hat{\mv v}$ may be obtained through a fixed aggregation function $\mathrm{agg}(\hat{\mv v}_1,\ldots, \hat{\mv v}_f)$ of the individual per-subframe decision. This function may only account for the most recent subframes to facilitate online decisions. An example is given by the majority rule that outputs the class selected over the majority of aggregated subframes.

\subsection{Training} \label{annt}
In order to optimize the model parameter vector $\mv \phi$ of the ANN, we assume access to a training data set $\mathcal{D}$ containing input-output pairs $(\mv u, \mv v)$, where $\mv u$ collects all the samples sensed by the $K$ devices and $\mv v$ is the corresponding target one-hot variable. The sensed signal $\mv u$ is formatted into $F$ subframes $\mv u_f$, with $f=1,\ldots,F$, as explained above. The loss between the probability vector $\hat{\mv v}_f=G_{\scriptsize \mv \phi}(\mv u_f)$ and the true target $\mv v$ is measured by the cross-entropy loss \cite{simeone2022machine}
\begin{align}
    L (\hat{\mv v}_f, \mv v)=-\sum_{j=1}^{D_v}v_j\log(\hat{v}_{j,f}),
	\label{cceloss}
\end{align}
where $v_j$ and $\hat{v}_{j,f}$ are the $j$th entry of vector $\mv v$ and $\hat{\mv v}_f$, respectively. We wish to select the model parameter vector $\mv \phi$ based on the training example $(\mv u, \mv v)$ by addressing the empirical risk minimization (ERM) problem
\begin{align}
	\min_{\scriptsize \mv \phi}~\sum_{\scriptsize (\mv u, \mv v)\in \mathcal{D}} \sum_{f=1}^F L(G_{\scriptsize \mv \phi}(\mv u_f), \mv v).
    \label{mcproblem}
\end{align}
Accordingly, the model parameters $\mv \phi$ can be updated based on stochastic gradient descent (SGD)-based rule
\begin{align}
	\mv \phi=\mv \phi-\eta \sum_{\scriptsize (\mv u, \mv v)\in \mathcal{B}} \sum_{f=1}^F \nabla_{\scriptsize \mv \phi} L(G_{\scriptsize \mv \phi}(\mv u_f), \mv v),
	\label{cgr}
\end{align}
where $\mathcal{B}\subseteq\mathcal{D}$ is a mini-batch of data points and $\eta$ is the learning rate. 

\section{NeuroComm} \label{neuro}

In this section, we present the proposed end-to-end neuromorphic architecture for remote inference over wireless channels, which we refer to as NeuroComm. As introduced in Sec. \ref{sec:intro} and detailed in Fig. \ref{model}, in NeuroComm, neuromorphic transmission and processing blocks are integrated with neuromorphic sensing. We proceed by describing neuromorphic encoding, decoding, and processing. Throughout this section, we consider the encoding and decoding SNNs to be fixed, i.e., to have fixed synaptic weights. In the next section, we will introduce the proposed hypernetwork-based architecture that enables optimization of encoding and decoding SNNs, as well as adaptation of the decoding SNN to current channel conditions.

\subsection{Neuromorphic Encoding}
With neuromorphic encoding, at each device $k$, the $D_u$ samples $\mv u_l^k$ produced by the sensor are encoded in an online, streaming, fashion by an SNN, across time instants $l=1,2,\ldots, L$. The samples output by each encoding SNN are fed to the impulse radio transmitting module. The encoding SNNs hence implement \emph{joint source-channel coding}, which is unlike the conventional solution reviewed in the previous section.

We investigate two distinct ways to implement neuromorphic encoding: one based on conventional \emph{time hopping} (TH) for impulse radio (see, e.g., \cite{win1998impulse}), and the other on an optimized solution that we refer to as \emph{learned time hopping} (LTH) modulation. As illustrated in Fig. \ref{TH}, the distinction between the two schemes is in the way in which bandwidth expansion is handled. In TH, the $k$th encoding SNN produces a single sample $\mv x_l^k$ for each sensed sample $\mv u_l^k$. Bandwidth expansion is then carried out via TH in order to produce the samples $\mv s_i^k$ to be transmitted on the wireless channel \eqref{y_l}. Specifically, as detailed below, TH outputs $L_b$ samples $\mv s_i^k$, with $i \in \mathcal{I}_l=\{(l-1)L_b+1,\ldots, lL_b\}$, for each sample by randomizing the timing of any spike in $\mv x_l^k$. In contrast, in LTH, the encoding SNNs directly produce the $L_b$ samples $\mv s_i^k$, with $i\in \mathcal{I}_l$, given each sensed sample $\mv u_l^k$.

For both TH and LTH, the encoding SNN at each device $k$, with $k=1,\ldots,K$, defines a causal mapping between the sensed time sequence $\mv u_l^k$, and the encoded time sequence. The encoded time sequence -- $\mv x_l^k$ for TH and $\mv s_i^k$ for LTH -- consists of $D_x=N^T$ spiking signals. Each of the $D_x$ signals is mapped to the output of one of the $N^T$ transmit antennas. We note that one could accommodate a larger number of antennas with $N^T>D_x$ by applying space-time codes for pulse-position modulation \cite{han2015multi}, but we do not further pursue this direction in this work.

\subsubsection{TH Modulation} 
\begin{figure}[t!]
	\begin{center}
		\subfigure[TH modulation.{\label{rep1}}]{\includegraphics[width=4.2in]{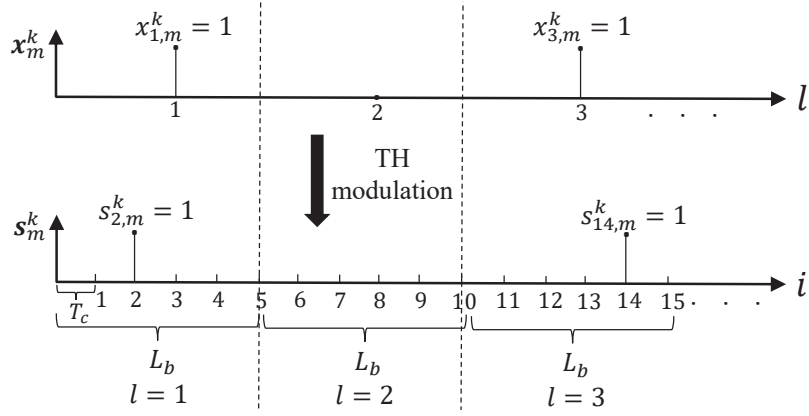}}\vfill
		\subfigure[LTH modulation.{\label{rep2}}]{\includegraphics[width=4.2in]{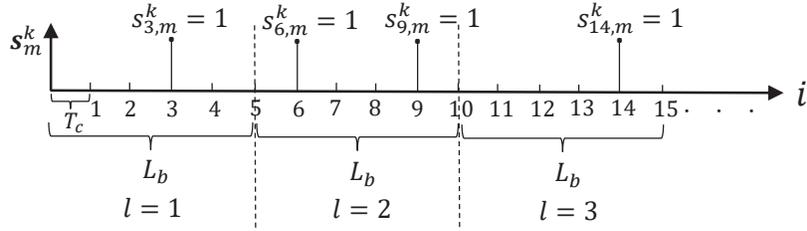}}
	\end{center}
	\caption{(a) \emph{Time hopping (TH) modulation}: Each output sample $x_{l,m}^k$ produced by the encoding SNN for the $m$th antenna at device $k$ is expanded to $L_b=5$ samples via TH. Accordingly, a spike, i.e., a sample $x_{l,m}^k=1$, is encoded as a single spike with a random time within the interval of $L_b$ samples corresponding to the $l$th sensing time step. (b) \emph{Learned time hopping (LTH) modulation}: The encoding SNN directly outputs $L_b$ samples $s_{i,m}^k$ with $i\in\mathcal{I}_l=\{(l-1)L_b+1,\ldots, lL_b\}$, for each sensed sample $u_{l}^k$. \label{TH}}
\end{figure}

As illustrated in Fig.~\ref{TH}-(a), with standard TH, the encoding SNN produces an $N^T \times 1$ vector of samples  $x_{l,m}^k$ for $m=1,...,N^T$ given each sensed sample $u_{l}^{k}$. Each sample $x_{l,m}^k$ is then expanded into $L_b$ samples $s^k_{i,m}$, for $i\in\mathcal{I}_l$, by randomly adding a discrete-time offset $c^k_{l,m}\in\{1,\cdots,L_b\}$ to the timing of each spike. For example, the spike $x_{1,m}^k=1$ in Fig.~\ref{TH}-(a) hops to position $c^k_{1,m}=2$ in the first interval of sensing time. The sequence of random hop times $c^k_{l,m}\in\{1,\cdots,L_b\}$ is not assumed to be known to the receiver. 

\subsubsection{LTH Modulation}

TH modulation is motivated by an attempt to avoid interference between waveforms transmitted by different antennas and devices by randomizing the position of a spike within a given interval of $L_b$ samples. However, in NeuroComm the goal is not to recover the individual transmitted signals, but rather to carry out an inference task. Therefore, TH may not be the most effective solution. Based on this observation, we introduce LTH, whereby we let the SNN utilize the available $L_b$ channel uses per sensed sample in a manner that is optimized (as we will study in the next section) for the given end-to-end task.

To this end, we expand the sensed signal $\mv u^k_l$ for each time instant $l$ by adding $L_b-1$ zero samples before feeding the sequence of samples to the encoding SNN. As a result, the encoding SNN directly produces $L_b$ output samples $s_{i,m}^k$, with $i\in \mathcal{I}_l$, in each interval of sensing time $l$. These samples are directly transmitted on the channel \eqref{y_l} as discussed next.
 
\subsection{Impulse Radio Transmission}

\begin{figure}[t!]
	\centering
	\includegraphics[width=6.2in]{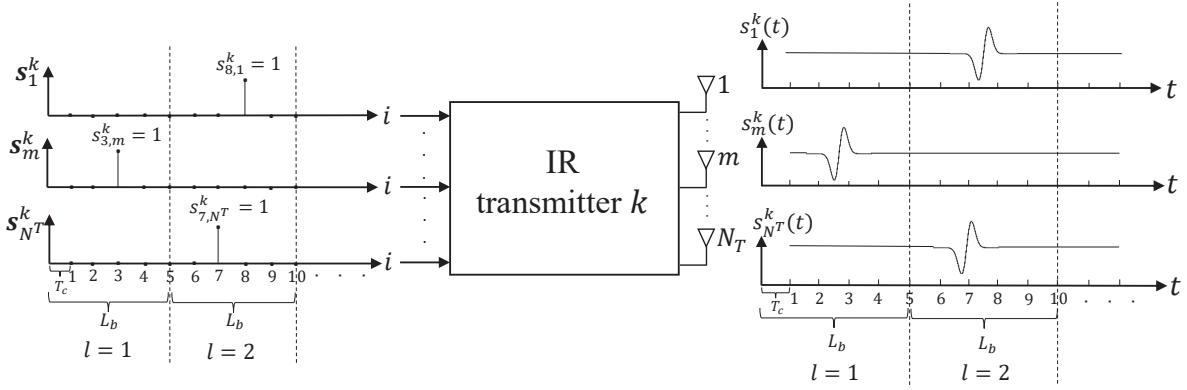}
	\caption{The impulse radio block takes as input the spiking signals produced by the encoding SNN, and possibly by TH modulation, and it produces the modulated signals $s_m^k(t)$, for $m=1,\ldots, N^T$.}
	\label{modulation}
\end{figure}

The samples $s_{i,m}^k$ produced by the encoding SNN are modulated onto antenna $m$ by device $k$ using pulse modulation. Accordingly, each antenna outputs a stream of pulses as illustrated in Fig. \ref{modulation}, and, unlike the conventional solution described in the previous section, information in NeuroComm is directly encoded in the timing of the spikes. Importantly, if the sequence $s_{i,m}^k$ is sparse, so is the signal transmitted by antenna $m$ at device $k$. This way, NeuroComm naturally controls the transmitted energy as a function of the activity in the sensed signals. If the sensors produce no spikes, the encoding SNN is also idle, implying that minimal computing energy and no transmission energy are consumed. In this regard, we also observe that, with TH, at most a single pulse is produced in each period of $L_b T_c$ seconds, while with LTH one may have multiple pulses, up to $L_b$, in each such period.

\subsection{Neuromorphic Decoding} \label{nde}
The modulated signals from the $K$ devices are transmitted over the channel, and the discrete-time received signals \eqref{y_l} are input to the decoding SNN, which operates at the same rate as the neuromorphic sensors. To feed the decoding SNN, we collect the received signals $\{y_{i,n}\}$ corresponding to the sensed time instant $l$, i.e., samples $y_{i,n}$ with $i\in\mathcal{I}_l$, into a $L_{\rm b}\times 1$ vector $\mv y_{l,n}$. Since this vector is generally complex, we define the $2L_{b}\times1$ vector 
\begin{align}
\bar{\mv y}_{l,n}=[\mathfrak{R}(\mv y_{l,n})^T, ~\mathfrak{S}(\mv y_{l,n})^T]^T, \label{spikingy}
\end{align}
with $\mathfrak{R}(\cdot)$ and $\mathfrak{S}(\cdot)$ being the element-wise real and imaginary parts of the input vector, respectively. The decoding SNN implements a causal mapping between the sequence of $2 L_b N^R \times 1$ vectors $\bar{\mv y}_l=[\bar{\mv y}_{l,1}^T,\ldots,\bar{\mv y}^T_{l,N^R}]^T$ and spiking signals $\bar{\mv v}_l\in\{0,1\}^{D_v}$ across time instants $l=1,\ldots,L$. Finally, the $D_v$ spiking signals in vector sequence $\bar{\mv v}_l$ are used as input to a block that makes the final inference decision.

For a classification task with $D_v$ classes, at each discrete time $l$, a decision can be made by decoding the output signal $[\bar{\mv v}_1,\ldots,\bar{\mv v}_l]$ produced thus far by the decoding SNN to a class index using standard methods, such as rate decoding (see, e.g., \cite{jang2019introduction}). With rate decoding, the $D_v$ signals $[\bar{\mv v}_1,\ldots,\bar{\mv v}_l]$ are first converted into a $D_v \times1$ vector by counting the spikes in each of the $D_v$ spiking signals, and then the $D_v \times 1$ vector of rates is passed through a softmax function. This can be written as 
\begin{align}
\hat{\mv v}_l=\textrm{softmax}(\bar{\mv V}_l \times \mathbf{1}), \label{soft}
\end{align}
where $\bar{\mv V}_l$ is the $D_v \times l$ matrix $\bar{\mv V}_l=[\bar{\mv v}_1,\ldots,\bar{\mv v}_l]$; $\mathbf{1}$ is the all-ones $l \times 1$ vector; and the softmax function outputs a $D_v \times 1$ vector with the $j$th entry given by $\textrm{softmax}_j(\mv x)=e^{x_j}/{(\sum_{j=1}^{D_v}e^{x_j})}$ for a $D_v \times 1$ input vector $\mv x=[x_1,\ldots,x_{D_v}]$.
Then, the class corresponding to the largest number of spikes across $L$ time steps, that is, to the largest entry in vector $\hat{\mv v}$, is selected.

\subsection{Spiking Neural Network Model}
\label{sec:snn-model}
In this work, each encoding/decoding SNN has a layered structure with fully connected layers. We describe here the general architecture using a  notation that does not specify the identity of the SNN, with the understand that superscripts can be added to distinguish encoding and decoding SNNs. For instance, we will write as $N_{\rm L}$ the number of layers with $N_{\rm L}^k$ being the number of layer for the $k$th encoding SNN and $N_{\rm L}^R$ being the number of layer for the decoding SNN. 

Each layer $\ell$, with $\ell=1, \ldots, N_{\rm L}$, consists of $N_\ell$ spiking neurons, and we denote by $\mv W_\ell$ the $N_{\ell+1}\times N_\ell$ weight matrix describing the connections between neurons in layer $\ell$ and layer $\ell+1$. Neurons follow the standard spike response model (SRM) \cite{skatchkovsky2021spiking}. Each spiking neuron $k$ outputs a binary signal $b_{k,l}\in\{0,1\}$, with ``1'' representing the firing of the spike and ``0'' an idle neuron at each time step $l=1,\ldots,L$.  Each neuron $k$ in layer $\ell+1$ receives inputs from the set $\mathcal{N}_\ell$ of neurons in layer $\ell$.

Following the SRM, each neuron $k$ maintains an internal analog state variable $o_{k,l}$, known as the \emph{membrane potential}, over time step $l$. Mathematically, the membrane potential $o_{k,l}$ is defined by the sum of filtered contributions from incoming spikes and from the neuron $k$'s own past outputs. Mathematically, we write the evolution of the membrane potential as 
\begin{align}
	o_{k,l}=\sum_{j\in\mathcal{N}_{\ell}}w_{k,j,\ell}\cdot(\alpha_l * b_{j,l})+\beta_l * b_{k,l},
	\label{potential}
\end{align}
where $w_{k,j,\ell}$ is the element $(k,j)$ of matrix $\mv W_\ell$, which corresponds to the synaptic weight between neuron $j\in\mathcal{N}_{\ell}$ and neuron $k$ in layer $\ell+1$; $\alpha_l$ represents the synaptic response to a spike from the presynaptic neurons $j\in\mathcal{N}_{\ell}$ to a postsynaptic neuron $k$; $\beta_l$ describes the synaptic response to the spike emitted by the neuron itself; and ``$*$'' is the convolution operator. Typical choices for synaptic spike responses include the first-order feedback filter $\beta_l=\exp(-l/\tau_{\rm ref})$, and the second-order synaptic filer $\alpha_l=\exp(-l/{\tau_{\rm mem}})-\exp(-l/{\tau_{\rm syn}})$, for $l=1,2,\ldots$, with finite positive constants $\tau_{\rm ref}$, $\tau_{\rm mem}$ and $\tau_{\rm syn}$.

Neuron $k$ outputs a spike at time step $l$ when its membrane potential $o_{k,l}$ passes some fixed threshold $\vartheta$, i.e.,
 \begin{align}
	b_{k,l}=\Theta(o_{k,l}-\vartheta),
	\label{spike}
\end{align}
where $\Theta(\cdot)$ is the Heaviside step function.

\section{Enabling Adaptation via Hyper-NeuroComm} \label{sec:training}
Deploying NeuroComm requires the specification of the synaptic weights $\mv W^1$, $\ldots$, $\mv W^K$ of the $K$ encoding SNNs, as well as of the weights $\mv W^R$ of the decoding SNN. In this section, we
introduce an end-to-end design strategy for all the synaptic weights. Design is carried out offline based on the availability of channel models or channel data. Importantly, the proposed approach does not require retraining for each channel condition. Rather, optimization of the synaptic weights is done only once, but the system is still able to adjust to changing channel conditions. Specifically, in the proposed architecture, the encoding SNNs’ weights are fixed irrespective of the current channel conditions, which are assumed to be unknown to the transmitters. In contrast, we allow for the weights $\mv W^R$ of the decoding SNN to adapt to the current channel condition by leveraging the received pilots (see Section II).

To this end, we propose a hybrid receiver architecture, termed Hyper-NeuroComm, which includes a conventional hypernetwork. As illustrated in Fig.~7, the hypernetwork takes as input the pilots received in the current frame, and outputs information controlling the weights of the decoding SNN. During offline training, the transmitters' weights, as well as the receiver's adaptation procedure via the hypernetwork, are jointly optimized. In this section, we first describe the Hyper-NeuroComm architecture, and then define the end-to-end training problem for the design of the encoding SNNs, decoding SNN, and hypernetwork. 

\subsection{Hyper-NeuroComm}
\begin{figure}[t!]
	\centering
	\includegraphics[width=6.4in]{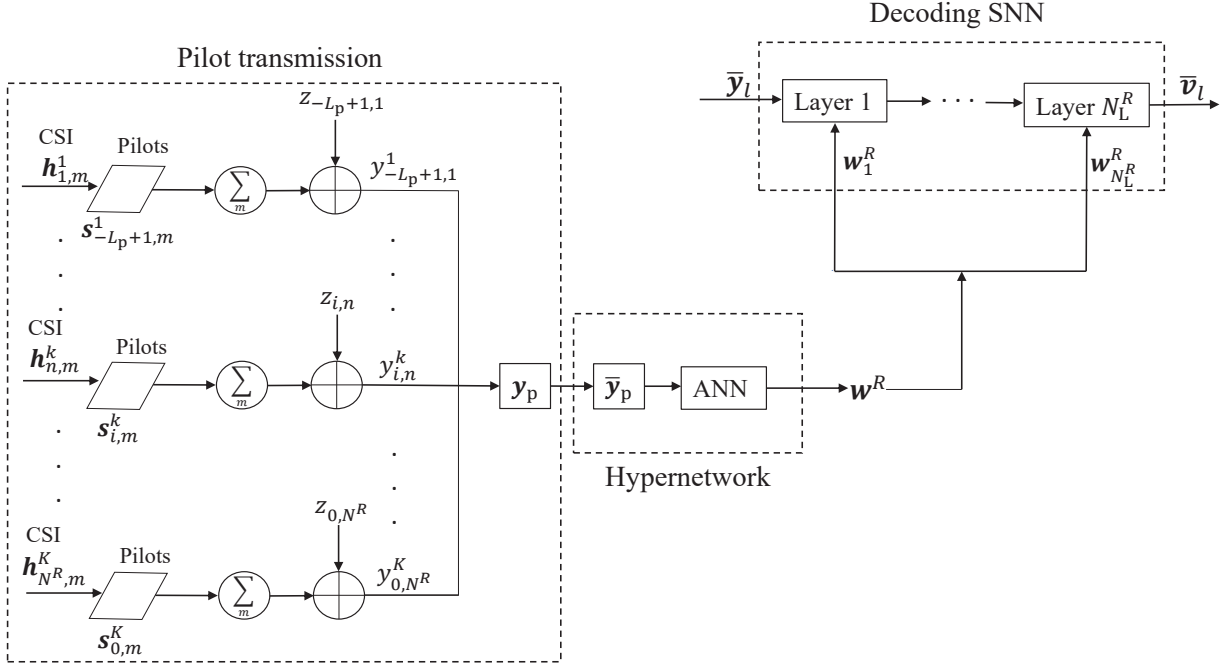}
	\caption{The Hyper-NeuroComm architecture, enabling the adaptation of the weights of the decoding SNN.}
	\label{hyper}
\end{figure}

Hypernetworks are neural networks whose outputs determine the weights of another neural network \cite{ha2016hypernetworks}. The target network in our setting is the decoding SNN. Accordingly, the ideal goal of the hypernetwork is to implement a parametric function 
 \begin{align}
	\mv W^R=\mathit{h}^{\rm id}_{\scriptsize \mv W^H}(\bar{\mv y}_{\rm p})
	\label{hyperf}
\end{align}
that maps the received pilots signals $\bar{\mv y}_{\rm p}$ in \eqref{y_p} to the weights $\mv W^R$ of the decoding SNN. The mapping \eqref{hyperf} is parametrized by the weights $\mv W^H$ of the hypernetwork. Via \eqref{hyperf}, the weights $\mv W^R$ of the decoding SNN can be adapted to the current channel conditions thanks to the received pilots $\bar{\mv y}_{\rm p}$. More practically, following \cite{liu2021hyperrnn}, we decrease the dimension of the output \eqref{hyperf} of the hypernetwork by producing a common scaling factor for each neuron, rather than a different weight for each synapse as in \eqref{hyperf}. The overall architecture of the receiver is shown in Fig.~\ref{hyper}, and is detailed next.

We start by collecting all the received pilot signals into a vector $\mv y_{\rm p}=[y_{-L_{\rm p}+1,1}^1,\ldots, y_{i,n}^k, \ldots,\\ y_{0, N^R}^K]^T\in\mathbb{C}^{L_\textrm{p}KN^R\times 1}$, and take the $2L_{\rm p}KN^R \times 1$ vector $\bar{\mv y}_{\rm p}=[\mathfrak{R}(\mv y_{\rm p})^T, ~\mathfrak{S}(\mv y_{\rm p})^T]^T$ as input to the hypernetwork. The output of the hypernetwork consists of $N_{\rm L}^R$ vectors $\{\mv w^R_1, \ldots, \mv w^R_{N_{\rm L}^R}\}$, with each $N^R_{\ell}\times 1$ vector $\mv w^R_\ell$ defining the scaling of the weights applied by the $N_\ell^R$ neurons in the $\ell$th layer. Specifically, following \cite{liu2021hyperrnn}, the weight matrix $\mv W^R_\ell$ for layer $\ell$ in the decoding SNN is set as 
 \begin{align}
	\mv W^R_\ell=\tilde{\mv W}^R_\ell \cdot \textrm{diag}\{\mv w^R_\ell\},
	\label{wc}
\end{align}
for $\ell=1,\cdots, N^R_{\rm L}$, where $\textrm{diag}\{\mv w^R_\ell\}$ is the $N_{\ell}^R \times N_{\ell}^R$ diagonal matrix with main diagonal given by the elements of $N_{\ell}^R \times 1$ vector $\mv w^R_\ell$; and the $N_{\ell+1}^R \times N_{\ell}^R$ matrices $\tilde{\mv W}^R_\ell$ for all $\ell$ are subject to optimization during training and they are fixed at run time, as the encoding SNNs' weights $\mv W^1, \ldots, \mv W^K$ (see next subsection). We write $\tilde{\mv W}^R$ for the collection of all matrices $\{\tilde{\mv W}^R_\ell\}_{\ell=1}^{N_{\rm L}^R}$, and the overall mapping implemented by the hypernetwork as 
 \begin{align}
	\tilde{\mv W}^R=\mathit{h}_{\scriptsize \mv W^H}(\bar{\mv y}_{\rm p}),
	\label{hyperff}
\end{align}
where $\mv W^H$ are the weights of the hypernetwork.

\subsection{Training Problem}
In this subsection, we address the problem of optimizing the model parameters $\mv W^1, \ldots, \mv W^K$ of the encoding SNNs, along with the weights $\tilde{\mv W}^R$ in \eqref{wc} of the decoding SNN, the weights $\mv W^H$ of the hypernetwork in \eqref{hyperff}, and the pilot sequences $\mv s_{\rm p}^1, \ldots, \mv s_{\rm p}^K$. We write as $\mv \theta=[\mv W^1, \dots, \mv W^K, \\ \tilde{\mv W}^R, \mv W^H, \mv s_{\rm p}^1, \ldots, \mv s_{\rm p}^K]$ the set of all parameters under optimization. As in Section \ref{annt}, we assume the availability of a training data set $\mathcal{D}$ given by pairs of inputs $\mv u=(\mv u^1,\dots,\mv u^K)$ produced by all sensors across all $L$ time instants, and the corresponding target one-hot vector $\mv v$. As described in Section \ref{nde}, the decoding SNN outputs spiking signals that are converted into a $D_v \times 1$ probability vector $\hat{\mv v}_l$ via \eqref{soft}. The output $\hat{\mv v}_l$ depends on the input $\mv u$ through the encoding SNNs' weights $\mv W^1,\ldots, \mv W^K$, the decoding SNN's weights $\mv W^R$, the channels $\mv h=\{\{\{\mv h_{n,m}^k\}_{n=1}^{N^R}\}_{m=1}^{N^R}\}_{k=1}^K$, as well as the additive channel noise in \eqref{y_l}. 

Using the cross-entropy loss \eqref{cceloss}, the training objective for Hyper-NeuroComm is defined as 
\begin{align}
	\min_{\scriptsize \mv \theta}~\sum_{\scriptsize(\mv u,\mv v)\in\mathcal{D}}\mathbb{E}\big[L\big(\hat{\mv v}_L, \mv v\big)\big],
    \label{problem}
\end{align}
where $\mathbb{E}[\cdot]$ denotes the expectation operator over channels $\mv h$ and noise, and we recall that the decoding SNN's weights $\mv W^R$ depend on the hypernetwork's weights $\mv W^H$ as in  \eqref{wc}. In practice, the expectation in \eqref{problem} is approximated by sampling channel realizations $\mv h$ and channel noise.

Problem \eqref{problem} is addressed via stochatic gradient descent in a manner similar to \eqref{cgr}. In order to compute the gradient, we follow the surrogate gradient approach \cite{neftci2019surrogate} by replacing the Heaviside step function \eqref{spike} with a differentiable surrogate function, namely a sigmoid $\sigma(x)=(1+e^{-x})^{-1}$.

\section{Experiments}\label{exp}
In this section, we provide experimental results with the aim of comparing the performance of the proposed end-to-end neuromorphic solution -- Hyper-NeuroComm -- with more conventional methods. As we detail next, we study a remote inference setting in which devices are equipped with DVS cameras observing parts of a screen displaying a handwritten digit. The goal is to ensure that the receiver be able to quickly classify the displayed image. 

\begin{figure}[htp]
	\centering
	\includegraphics[width=6.2in]{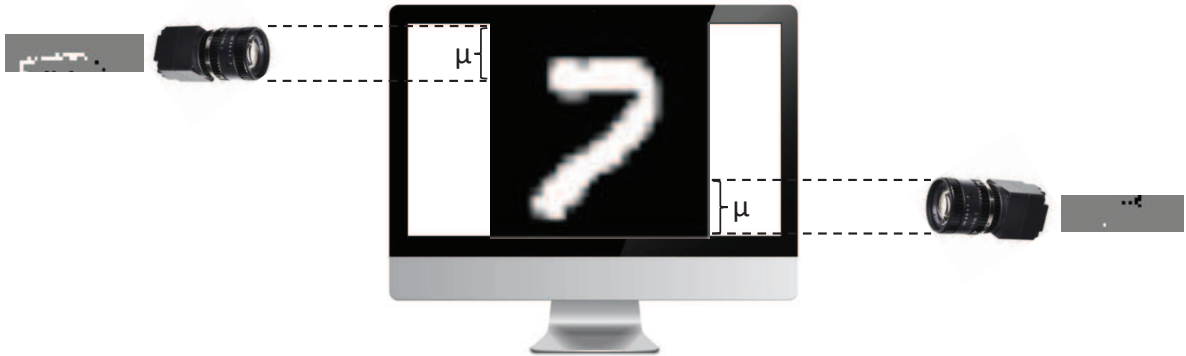}
	\caption{An illustration of a system with two devices $(K=2)$, each equipped with a DVS camera that observes a fraction $\mu$ of an image displayed on a screen. In the depicted example, the fraction $\mu$ is smaller than $0.5$,  and hence the two devices cannot collectively observe the whole image.}
	\label{simulation}
\end{figure} 

\subsection{Experimental Setting}
To set up the scenario, we consider the MNIST-DVS dataset \cite{serrano2015poker}, which contains labelled $26 \times 26$ spiking signals of duration $L=80$ samples. Each data points contains $26 \cdot 26 = 676$ spiking signals, which are recorded from a DVS camera that is shown moving handwritten digits $``1"$ and $``7"$ on a screen.  The data set contains $900$ examples from each class for training, and $100$ examples for testing. The neuromorphic sensor at each device records a fraction $\mu<1$ of the image. Accordingly, each device has an input signal consisting of $D_u = \lceil{\mu \cdot 676}\rceil$ spiking signals. Specifically, for the case of $K=2$ sensors, we assume, as shown in Fig.~\ref{simulation}, that the first device observes the fraction of top $\mu$ rows of the image, while the second device measures the bottom $\mu$-fraction of the rows. Note that when $\mu>1/2$ there is an overlap of size $2\mu-1$ between the subimages observed by the two sensors. 

Each sensor has $N^T=10$ antennas, and the decoder has $N^R=20$ antennas. The encoding SNNs has 128 hidden neurons, and $D_x =N^T= 10$ output neurons.  The decoder has $D_y = 20$ input neurons, 128 hidden neurons, and $D_v = 2$ output neurons, each corresponding to one of the two classes. The hypernetwork ANN is composed of one hidden layer with 1024 neurons. All layers are fully connected in the SNNs and in the hypernetwork. The pilots are initialized as white Gaussian noise and then subject to optimization during training by tackling problem \eqref{problem}. The number of pilots $L_{\rm p}$ is set to 64.

The transmitter uses Gaussian monopulses, and the receiver  performs classification at each discrete time instant. The number of paths $N_{P}$ is set to 5, all paths have the same average power $1/{N_P}$, and they are characterized by Rayleigh fading. The multipath delays are set to $0$, $T_c$, $2T_c$, $3T_c$ and $4T_c$. Unless otherwise stated, we consider the per-frame power constraint \eqref{pc}, and the average signal-to-noise ratio (SNR) for each device $k$ is defined as the average symbol power over noise power $\big( E^k_{\rm fr}/(D_x L L_b)\big)/N_0$. When considering per-symbol energy constraint \eqref{pc}, the SNR is defined as $E_{\rm s}^k/N_0$. The average SNR is set to $10$ dB. We train all the schemes based on the MNIST-DVS dataset with $1000$ channel and noise realizations for each sample, while the test results are averaged over $300$ channel realizations.
\subsection{Benchmarks}
    Given that this is the first work to propose the idea of neuromorphic computing for remote inference, there are no direct benchmarks involving neuromorphic technologies. Therefore, in order to benchmark the proposed approach, we rely on state-of-the-art conventional frame-based strategies that adopt traditional multi-access technologies and separate source and channel coding. We also consider several simplified variants of the proposed neuromorphic architecture.
    
For the proposed Hyper-NeuroComm, we implement both standard TH and LTH (see Sec. \ref{neuro}). Furthermore, we consider the following benchmarks.
\begin{itemize}
\item \emph{Frame-based digital  transmission:} Following the description in Section \ref{conventional method}, we implement frame-based digital transmission by following \cite{skatchkovsky2020end}. The transmitter collects $L_{\rm enc}=20$ time samples from the sensor to create a subframe. The subframe is encoded by source and channel encoding separately. Source encoding is performed using a Vector-Quantization Variational Autoencoder (VQ-VAE) \cite{van2017neural}, followed by LDPC channel encoding with code rate set to 2/3. The encoded data is transmitted subframe by subframe using BPSK to the receiver. Each subframe is decoded using the belief propagation algorithm, decompressed by the VQ-VAE, and then classified by an ANN, which consists of a single layer of 512 hidden neurons. Frame-based digital transmission is implemented with ALOHA to manage interference between the transmitted signals from these two devices. Accordingly, each device transmits a subframe with probability $0.5$, and subframe transmission is successful only when there are no concurrent transmissions. The aggregation function adopts a majority rule over the received subframes. The DNN is trained using SGD-based optimization based on the examples $(\mv u, \mv v)$ as per \eqref{cgr}.
\item \emph{NeuroComm with joint learning:} To set a lower bound on the performance of the proposed Hyper-NeuroComm architecture, we consider an alternative design method whereby the decoding SNN model parameters are not adapted to the channel realization using pilots. This solution is derived by removing the hypernetwork, and addressing problem \eqref{problem} directly over encoding and decoding SNN's weights.
\item \emph{NeuroComm with ideal per-channel learning:} To provide a performance upper bound, we also consider an ideal situation in which for each given channel realization we have abundant data to train the model parameter $\mv \theta$ to convergence. This is done by tackling problem \eqref{problem} via SGD by drawing a large number of channel and noise samples.
\end{itemize}

\subsection{Results: Single Device}
\begin{figure}[t!]
	\centering
	\includegraphics[width=4.5in]{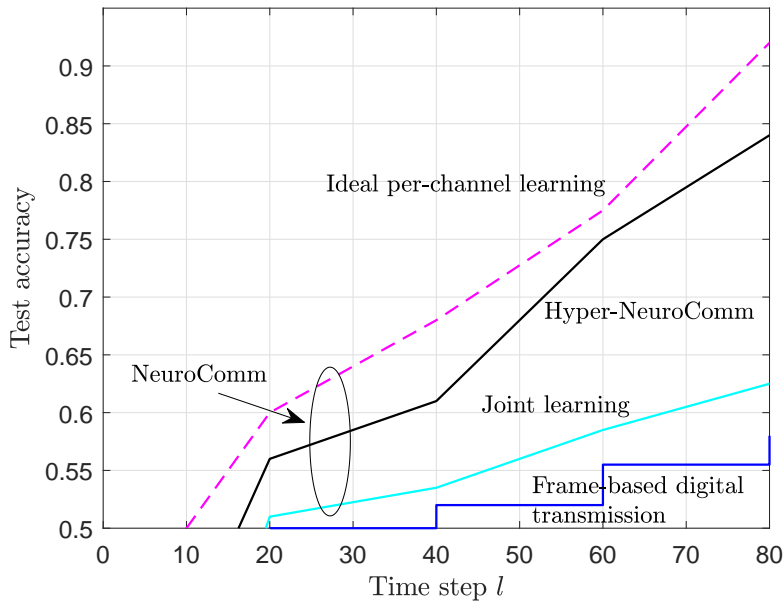}
	\caption{Test accuracy versus discrete time $l$ during transmission of an image with LTH $(K=1$, $\mu=1$, $L_b=6$, $SNR=10$ dB$)$.}
	\label{T}
\end{figure}

\begin{figure}[t!]
	\centering
	\includegraphics[width=4.5in]{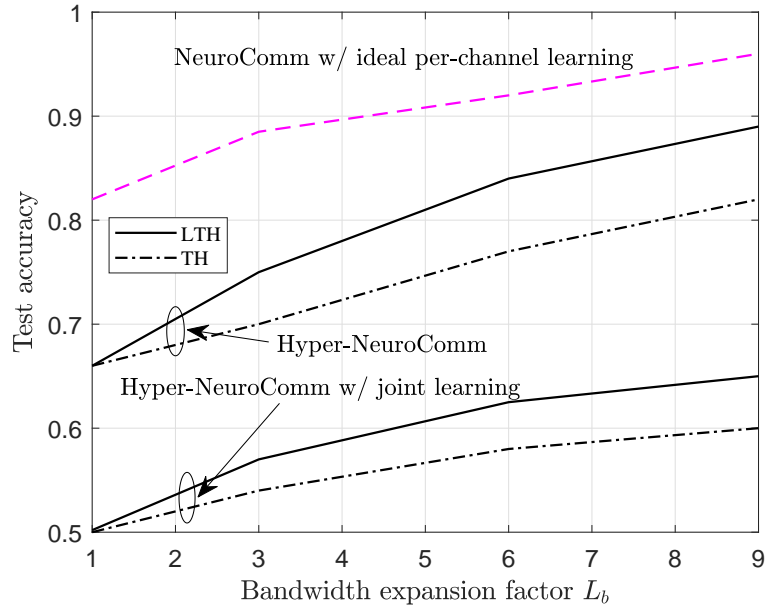}
	\caption{Test accuracy versus $L_b$ for NeuroComm $(K=1$, $\mu=1$, $SNR=10$ dB$)$. }
	\label{Lf}
\end{figure}

 We first consider a baseline scenario in which there is only one device observing he entire image $\mv u$, i.e., we have $K=1$ and $\mu=1$. We recall that reference \cite{skatchkovsky2020end} also considered the scenario with $K=1$ and $\mu=1$, but only under a simple additive Gaussian noise channel with no fading. Therefore, the performance analysis reported here, which accounts for the impact of fading and the corresponding novel role played by the pilots and by the hypernetwork, is novel. We first focus on LTH in Fig. \ref{T}, which generally outperform TH, and then compare LTH and TH in Fig. \ref{Lf}.

In Fig.~\ref{T}, we evaluate the evolution of the test accuracy during the observation of the signals from the sensors. The horizontal axis reports the discrete time $l$ elapsed since the start of the observation of the image by the sensors. The test accuracy of all schemes is seen to improve over time, as they can make decision based on more samples. However, the frame-based approach enables an improvement of the accuracy level only after each subframe is received, decoded, and processed by the DNN. NeuroComm with joint learning performs slightly better than frame-based transmission, offering a gradual increase in test accuracy. However, the gain is limited by the lack of adaptation to changing conditions.

Hyper-NeuroComm is seen to outperform NeuroComm with joint learning, achieving a test accuracy close to the ideal case of per-channel learning. This demonstrates the advantages of the proposed hypernetwork-based architecture in transferring useful knowledge across fading realizations, while also being able to adapt to current channel conditions. 

\begin{figure}[t!]
	\centering
	\includegraphics[width=4.5in]{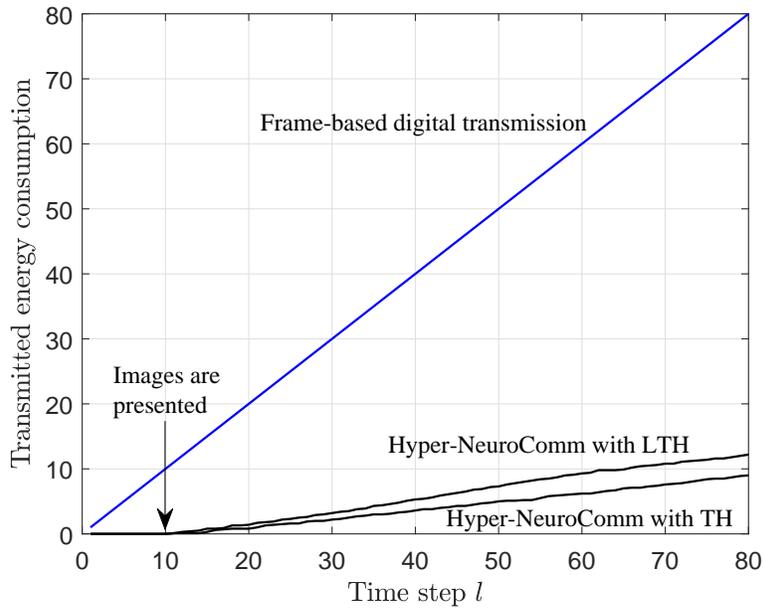}
	\caption{Cumulative transmitted energy consumption versus time step $l$ under a per-symbol energy constraint $(K=1$, $\mu=1$, $L_b=5$, $SNR=10$ dB$)$. }
	\label{energy}
\end{figure} 

Fig.~\ref{Lf} shows the test accuracy as a function of the bandwidth expansion factor $L_b$. We recall that a larger value of $L_b$ provides the encoding SNN with a proportionally larger interval of time over which to encode the sensed input. The test accuracy is observed to generally increases with the bandwidth expansion $L_b$, owing to the decreased interference between the transmitted streams from the $N^T$ antennas. Furthermore, LTH is seen to be effective in making a more efficient use of the increased encoding period. Rather than only managing interference between streams, LTH can in fact effectively use the larger number of encoded symbols per sensed sample to enhance the efficiency of joint source-channel encoding from sensed to transmitted signals.

Finally, Fig.~\ref{energy} demonstrates the cumulative transmitted energy consumption as a function of the time step $l$ for frame-based digital transmission and Hyper-NeuroComm. Unlike the previous figures, we adopt the per-symbol energy constraint \eqref{sc}. The images are presented starting from time step $10$. While frame-based digital transmission scheme consumes constant power, NeuroComm only consumes energy when images are presented to the DVS cameras. Furthermore, the cumulative energy consumption is lower than for frame-based transmission, due to the sparsity of the spiking signals output by the sensors and by the encoding SNNs. Finally, the LTH scheme consumes more energy than TH, since the LTH may produce multiple pulses during each time step. 

\subsection{Results: Multiple Devices}
\begin{figure}[t!]
	\centering
	\includegraphics[width=4.5in]{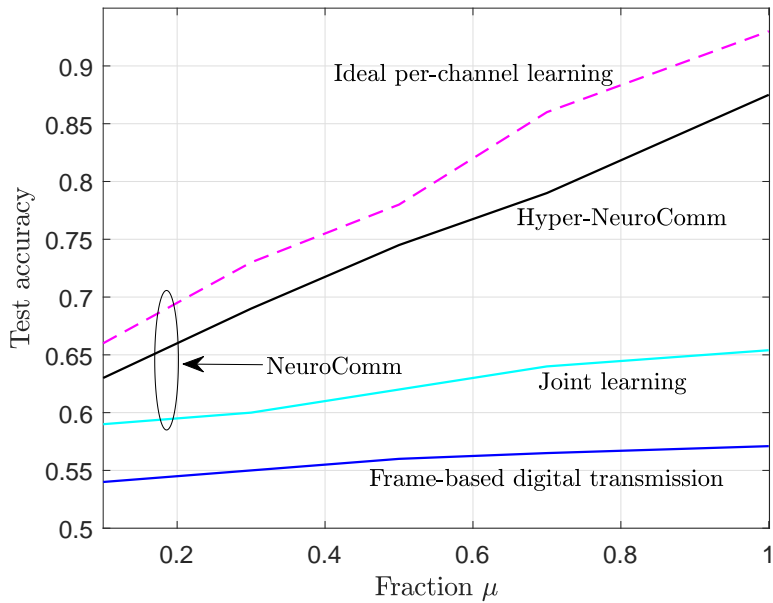}
	\caption{Test accuracy versus fraction $\mu$ of observed image during transmission of an image with LTH $(K=2$, $SNR=10$ dB$)$.}
	\label{mu}
\end{figure}

We now move on to a setting with $K=2$ devices, each observing a fraction $\mu$ of the screen as described in Fig. \ref{simulation}. In Fig. \ref{mu}, we shows the test accuracy as a function of the fraction $\mu$ of the sensed signals under LTH modulation. Test accuracy is seen to improve as sensors record a larger fraction of an image. However, only Hyper-NeuroComm is able to fully benefit from the more informative sensed signals, achieving a performance close to ideal per-channel learning.

\begin{figure}[t!]
	\centering
	\includegraphics[width=4.5in]{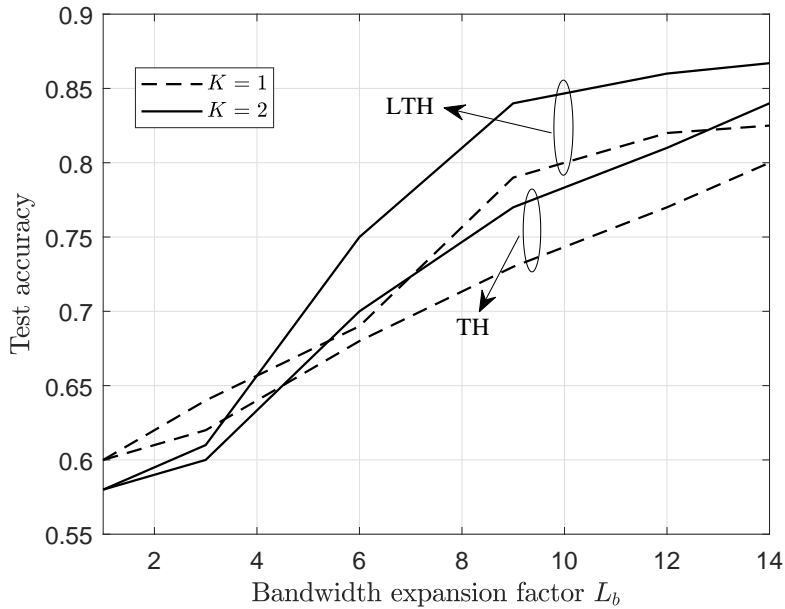}
	\caption{Test accuracy versus $L_b$ for Hyper-NeuroComm $(\mu=0.75$, $SNR=10$ dB$)$.}
	\label{Lff}
\end{figure} 

Fig.~\ref{Lff} plots the test accuracy versus the bandwidth expansion factor $L_b$ for Hyper-NeuroComm with $K=1$ and $K=2$ devices under LTH and TH. An interesting question arising from this experiment is: Under which conditions does having two sensors help? The figure shows that, if the bandwidth expansion is too low, e.g., if $L_b<4$ for LTH, it is preferable to have a single device active, despite the fact that the device only observes a fraction $\mu$ of the entire image. Furthermore, it can be noted that, while in the single-device case Hyper-NeuroComm with LTH achieves its best performance with a smaller value of $L_b$, e.g., $L_b=9$, a larger $L_b$ is called for in order to mitigate interference from multiple devices.

\section{Conclusion}\label{con}
This paper has introduced NeuroComm, a novel neuromorphic end-to-end solution for remote wireless inference in applications involving always-on online monitoring. The proposed system integrates neuromorphic sensing and neuromorphic computing with impulse radio transmission, enabling the adaptation of energy consumption to the dynamics of the monitored environment while ensuring low-latency inference. At a system level, the innovations of this paper also include the use of a hypernetwork at the receiver side to optimize adaptation of the receiver to the current channel conditions. Extensive experimental results have demonstrated the benefits of neuromorphic sensing and computing for the implementation of wireless cognition in terms of time-to-accuracy and energy consumption. It is the hope of the authors that this work will pave the way for further study of the synergy between neuromorphic sensing and processing technologies with wireless communications. A particularly interesting avenue for research would be the development of a testbed incorporating software-defined radio that implement IR, as well as neuromorphic hardware, possibly accessed using cloud services such as Intel's Lava \cite{lava}. This would make it possible to assess the relative contributions of processing and transmit-receive chains on the overall energy consumption. Furthermore, one could consider other types of inference tasks such as tracking \cite{hu2016dvs}. Finally, the IR interface could be leveraged for conventional data transmission as well as a probing signal for sensing applications \cite{nisac}.

\small{
\bibliographystyle{IEEEtran}
\bibliography{references}}
\end{document}